\documentclass[aps,prb,twocolumn,showpacs,showkeys,superscriptaddress,floatfix]{revtex4}%preprint,
\usepackage{amsmath,graphicx,color}
\usepackage{graphicx}
\usepackage{amsmath}
\usepackage{amssymb}
\usepackage[T1]{fontenc}
\usepackage{amsmath,graphicx,color}

\usepackage{amsmath,graphicx}

\begin{document}
\title{Antiferromagnetism in hexagonal graphene structures: rings vs dots}
\author{M. Gruji\'c}\email{marko.grujic@etf.bg.ac.rs}
\affiliation{School of Electrical Engineering, University of Belgrade, P.O. Box 3554,
11120 Belgrade, Serbia}
\author{M. Tadi\'c}\email{milan.tadic@etf.bg.ac.rs}
\affiliation{School of Electrical Engineering, University of Belgrade, P.O. Box 3554,
11120 Belgrade, Serbia}
\author{F. M. Peeters}\email{francois.peeters@ua.ac.be}
\affiliation{Department of Physics,  University of Antwerp,
Groenenborgerlaan 171, B-2020 Antwerp, Belgium}
\begin{abstract}
The mean-field Hubbard model is used to investigate the formation
of the antiferromagnetic phase in hexagonal graphene rings with
inner zigzag edges. The outer edge of the ring was taken to be
either zigzag or armchair, and we found that both types of
structures can have a larger antiferromagnetic interaction as
compared with hexagonal dots. This difference could be partially
ascribed to the larger number of zigzag edges per unit area in rings
than in dots. Furthermore, edge states localized on the inner ring
edge are found to hybridize differently than the edge states of
dots, which results in important differences in the magnetism of
graphene rings and dots. The largest staggered magnetization is found
when the outer edge has a zigzag shape. However, narrow rings with
armchair outer edge are found to have larger staggered
magnetization than zigzag hexagons. The edge defects are shown to
have the least effect on magnetization when the outer ring
edge is armchair shaped.
\end{abstract}
\maketitle

Recent tremendous progress in graphene research is driven by its
remarkable properties, e.g. high crystalline quality, high
electron mobility, lack of a band gap, and a minimal possible
thickness, to name a few.\cite{nov04} The mentioned properties are
advantageous for various applications of graphene, such as
piezoelectric devices,~\cite{ong12}
supercapacitors,~\cite{el-kady12} photodetectors,~\cite{photod12}
and field-effect transistors.~\cite{lin10,brit12} Furthermore, it
has been predicted that graphene structures could exhibit  {\it
magnetic ordering} which is potentially advantageous for
spintronic applications.~\cite{oleg10,son06} This effect is
essentially related to either the global or local imbalance of
sublattice atoms in bipartite lattices. An imbalance might give
rise to zero energy states in the electron spectrum. These states
are localized near the zigzag edges or vacancies, and along with
the repulsive electron-electron ({\it e}-{\it e}) interaction could eventually
lead to a spin polarization of the ground state of the
system.~\cite{oleg10} Furthermore, the spins on the same
sublattice are found to exhibit ferromagnetic coupling along the
graphene edges, whereas the spins on different sublattices along
the graphene edges couple antiferromagnetically.

In theory, magnetic ordering has been demonstrated for graphene
flakes,~\cite{rossier07} nanoribbons,~\cite{rossier08} and
vacancies in bulk graphene.~\cite{palac08} On the other hand,
experimental reports on magnetism in graphene structures are rare
and conflicting. They range from the detection of ferromagnetic or
antiferromagnetic ordering~\cite{matte09,enoki09,joly10} to
measurements of defect-induced paramagnetism.~\cite{ney11,nair12}
Magnetic ordering was even found to be preserved at room
temperature~\cite{wangliang09,hong12}. The essential cause of
magnetism in graphene is the existence of a peak in the density of
nonbonding edge states near the Fermi energy. However, due to the
high reactivity of these states, magnetism might be strongly
suppressed.~\cite{kunst11} Several theoretical studies offered
explanations for a diversity of phenomena related to magnetic
ordering and its suppression, which might occur by means of
nonmagnetic edge passivation, edge reconstruction, or vanishing
of spin correlations with increasing
temperature.~\cite{kunst11,yazyev08} Hence, in order to
experimentally detect magnetic ordering graphene samples should be
kept under rigorously controlled conditions. Yet, various
applications of this effect have been proposed. They involve
half-metallicity with electrically controlled spin
propagation,~\cite{son06} defect induced spin
filtering,~\cite{wimmer08} and spin logic
devices.~\cite{wang09,ezawa10}

In this report we employ the mean-field Hubbard model to study the
formation of local magnetic moments in hexagonal graphene rings.
Our aim is to explore how magnetic ordering is affected by the
ring size and the edge type. In order to identify different
hexagonal rings, we introduce the following notation which might
be visually aiding. We assume that the type of the inner ring edge
is zigzag, and that $N$ unit hexagons are adjacent to this
boundary. The outer ring edge is assumed to be comprised of either
$M$ dimers if it is of armchair type, or $M$ unit hexagons if it
is of zigzag type. Therefore, the ring is denoted by $M:N$. As an
example, consider the ring shown in Fig. 1, which  is assumed to
be formed out of the hexagonal dot with armchair edge, which
contains seven dimers at each side of the hexagon, as shown in Fig.
1(a). The ring is formed when the carbon atoms around the center
of the dot are removed, as depicted in Figs. 1(a) and (b).
Potentially these exotic structures could be manufactured via
substitutional doping of boron nitride nanostructures with
carbon.\cite{wei11} Because the edge of the removed dot has four unit
hexagons at each side, the ring is denoted as $7_{AC}:4_{ZG}$. The
distributions of the magnetic moments in the graphene rings will
be compared with the magnetic moment distributions in the
hexagonal graphene dots. Those dots are assumed to have zigzag
edges, and are labeled by $N_{ZG}$, where $N$ has the same meaning
as the symbol $M$ for the rings.

Magnetic ordering of a graphene structure is governed by Lieb's
theorem.~\cite{lieb89} It states that the total ground-state spin
of a bipartite lattice with repulsive {\it e}-{\it e} interaction as described
by the Hubbard model equals half of the difference of the
sublattice sites. For symmetrical structures, this rule is related
to the arrangement of the carbon atoms with respect to lines of
reflection symmetry in the graphene plane: if there is a symmetry
line which does not intersect any of the carbon atoms the total
ground state spin is zero; otherwise there exists a finite
magnetic moment. All the hexagonal rings analyzed here possess
such a symmetry, thus their total magnetization equals zero,
unlike triangular rings which display a ferrimagnetic
phase.~\cite{potasz11} However, Lieb's theorem does not dictate
the distribution of the local magnetic moments or the lack of zero
energy states. Furthermore, the number of zero-energy states in
the analyzed ring is an integer multiple of six, which is a
consequence of the $C_{6v}$ symmetry of the ring. In the
$9_{AC}:10_{ZG}$ ring, six zero energy states are found, which
agrees with graph theory, and which is a topological property
related to the nonperfect matching of the $p_z$
orbitals.~\cite{oleg10}

The Hubbard Hamiltonian
\begin{equation}
    H=H_0+H_I,
\end{equation}
is employed to compute the distribution of magnetic moments. $H_0$
is the noninteracting part, which represents the nearest neighbor
tight-binding Hamiltonian, and is given by
\begin{equation}
    H_0=-t\displaystyle\sum\limits_{<i,j>,\sigma}c_{i\sigma}^\dagger c_{j\sigma},
\label{h0}
\end{equation}
where $c_{j\sigma}$ and $c_{j\sigma}^\dagger$ are the annihilation
and creation operators, respectively, and $t$ denotes the hopping
integral. The interacting part $H_I$ describes the e-e interaction
\begin{equation}
    H_I=U\displaystyle\sum\limits_i\left(n_{i\uparrow}\langle
    n_{i\downarrow}\rangle+n_{i\downarrow}\langle
    n_{i\uparrow}\rangle-\langle n_{i\uparrow}\rangle\langle
    n_{i\downarrow}\rangle\right), \label{hi}
\end{equation}
where  $n_{i\sigma}=c_{i\sigma}^\dagger c_{i\sigma}$ is the number
operator, and $U$ denotes the on-site Coulomb repulsion energy for
each pair of electrons with the opposite spins orbiting the same
atom.\cite{futnota1} Equation~(\ref{hi}) is obtained within the
mean-field approximation, which assumes that the spin-up
(spin-down) electrons interact with the average density of
spin-down (spin-up) electrons on a particular atomic site.

In our calculations, we take $t=2.7$ eV and $U=1.2t$.\cite{oleg10}
We note that there is no consensus on the actual value of the
strength of the Coulomb interaction to be used in the Hubbard
model in graphene. Recent density functional theory (DFT) calculations came up with a value
closer to $U=3.4t$.\cite{wehling11} However, having in mind that
the mean-field approximation can overestimate the tendency for
magnetic order for large $U$,\cite{feldner10} we chose the more
conservative value of $U=1.2t$. The solution is then obtained by
means of a self-consistent procedure which starts from an initial
distribution of the spins, and ends when the maximum change of the
electron density over the atomic sites drops below $10^{-5}$. When
the self-consistent spin densities are determined, the magnetic
moment per site $m_i$ is computed as
\begin{equation}
    m_i=\langle s_i^z\rangle=\left(\langle
    n_{i\uparrow}\rangle-\langle n_{i\downarrow}\rangle\right)/2.
\end{equation}

For the antiferromagnetic order parameter we take the staggered
magnetization
\begin{equation}
\mu_s^z=\frac{1}{N}\sum_i(-1)^i\langle s_i^z\rangle,
\end{equation}
where $(-1)^i$ symbolizes that we sum up the contributions from
opposite sublattices with opposite signs. This is the appropriate
order parameter for antiferromagnetism when examining spin
polarization occurring in bipartite lattices. The larger $\mu_s^z$
the stronger is, the antiferromagnetic phase. In addition to
$\mu_s^z$, the shift in the electron and the hole energy spectra
which arises from the magnetic order is quantified as $\Delta
E=\left(E^{HOS}+E^{LUS}\right)/2$, where $E^{HOS}$ and $E^{LUS}$
are the highest occupied and lowest unoccupied states in the
ground state at half filling, respectively. Note that in the nonmagnetic state
we have $\Delta E=0$. We will explore how the maximum magnetic
moment $m_{max}$ varies with the ring width.

\begin{figure}
\centering
\includegraphics[width=8.6cm]{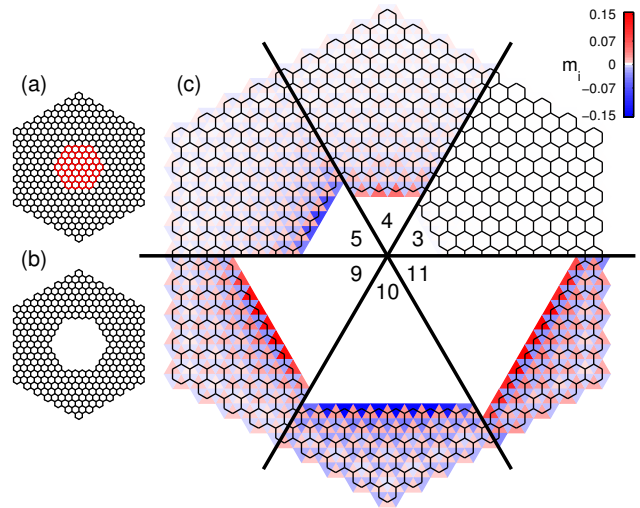}
\caption{(Color online) (a) The dot (red color) with four atoms at
the zigzag edge removed from the larger dot (black color) which
has seven dimers at the dot edge. (b) The formed ring is labeled by
$7_{AC}:4_{ZG}$. (c) The distribution of magnetic moments in the
$9_{AC}:N_{ZG}$ ring shown in a sextant of the ring for $N$ taking
the values 3, 4, 5, 9, 10 and 11. The majority spin is labeled by
both orientation and color of a triangle centered at an atomic
site. The local magnetic moment value is proportional to the color
intensity.}
\end{figure}

The distribution of the local magnetic moments in the
$9_{AC}:N_{ZG}$ rings for several values of $N$ is shown in
Fig.~1(c). The symmetry of each hexagonal ring is $C_{6v}$,
whereas the symmetry of the magnetic moment distribution is
$IC_{6v}$, i.e. the magnetic moments alter sign when rotated over
$\pi/3$ rad. Therefore, it suffices to display the distribution of
magnetic moments in sectors of $\pi/3$ rad, as done in Fig.~1(c),
which combines the sectors of different $N$. Orientation and color
of triangles denotes the orientation of the magnetic moments, and
the absolute value of $m_i$ is depicted by color intensity.

\begin{figure}
\centering
\includegraphics[width=8.6cm]{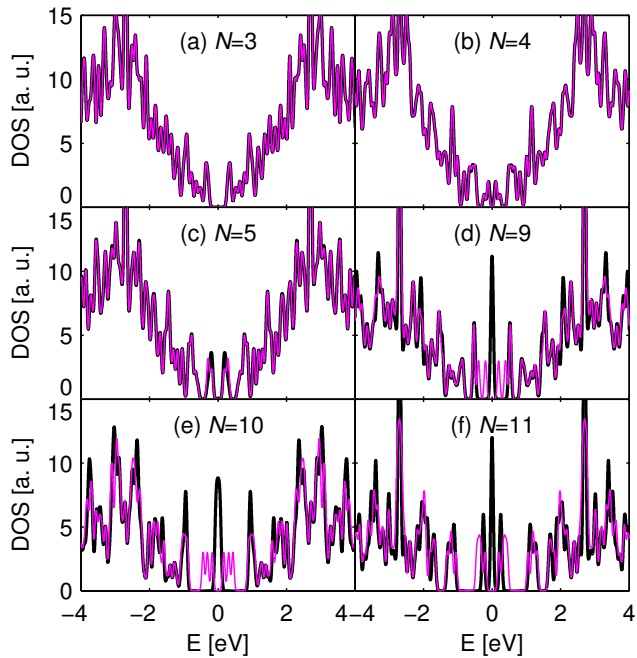}
\caption{(Color online) Density of states in the $9_{AC}:N_{ZG}$
rings for $N$ taking the values 3, 4, 5, 9, 10 and 11 in the
noninteracting system (black lines) and the interacting system
(purple lines).}
\end{figure}

It is evident in Fig.~1(c) that both the appearance of staggered
magnetization and the total magnetic moment situated on the inner
edge of the ring depend on the ring width. Furthermore, we
observe a phase change from nonmagnetic order for $N\leq 3$ to
antiferromagnetic order for $N\geq 4$, which is similar to
previous calculations for zigzag hexagonal graphene
dots.\cite{rossier07,rossier08} No magnetic ordering for zigzag
segments shorter than three unit cells is found because of the close
proximity of the opposite sublattice imbalance on the adjacent
sides of the ring inner edge. When this edge is short, the edge
states on the different sides of the inner ring boundary are
subject to strong hybridization, and therefore their energy is
lifted from the Dirac point. Hence, spontaneous spin polarization
does not occur, which is similar to the case of
nanoribbons.~\cite{dressel96}

\begin{figure}
\centering
\includegraphics[width=6.8182cm]{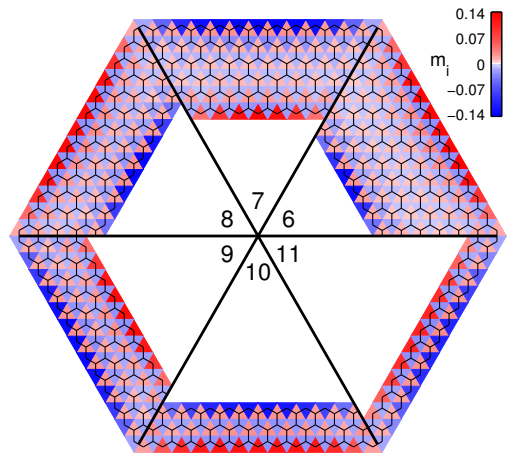}
\caption{(Color online) Distribution of magnetic moments in the $13_{ZG}:N_{ZG}$ rings for $N$ ranging from 6 to 11.}
\end{figure}

For $N\geq 4$, the spatial spin symmetry is broken due to the {\it e}-{\it e}
interaction. When the ring width decreases, the maximum magnetic
moment, which is located near the middle of the zigzag edge
segment increases. Furthermore, nonzero magnetization is build up
on the outer ring edge, and it increases when the ring width
decreases. However, as a consequence of the increasing influence
of the outer edge with decreasing ring width, the difference
between the distributions of the magnetic moments on the two edges
is not large for $N=10$ and $N=11$. Similarly, the staggered
magnetization increases when the ring width decreases.

Figure 2 shows how the density of states (DOS) of the
$9_{AC}:N_{ZG}$ rings (the cases depicted in Fig.~1) varies with
$N$. The density of states for the noninteracting (interacting)
case is displayed by the black (purple) lines. In order to align
the interacting and noninteracting spectra for easier comparison
we subtracted $\Delta E$ for each interacting spectrum. Note that
the density of states is spin degenerate, which is in accordance
with Lieb's theorem. For $N=3$, magnetic order is not present,
therefore the energy dependence of the density of states for the
interacting and noninteracting systems coincide [see Fig. 2 (a)].
The interacting and noninteracting electron case exhibit a small
difference in the energy dependence of the DOS for rings with
$N=4$ and $N=5$, which is shown in Figs. 2(b) and 2(c). As could be
inferred from Fig. 1(c), the magnetization along the inner ring
edge is rather small for these values of $N$. For larger $N$, the
discrepancy between the DOS's for the interacting and
non-interacting systems becomes larger, as demonstrated by
Figs.~2(d)-(f) for $N=9$, $10$, and $11$. In all these cases,
appreciable DOS for the noninteracting system is found around
zero energy. Such a configuration becomes unstable in the presence
of {\it e}-{\it e} interactions, which results in the appearance of an
interaction gap.

\begin{figure}
\centering
\includegraphics[width=8.6cm]{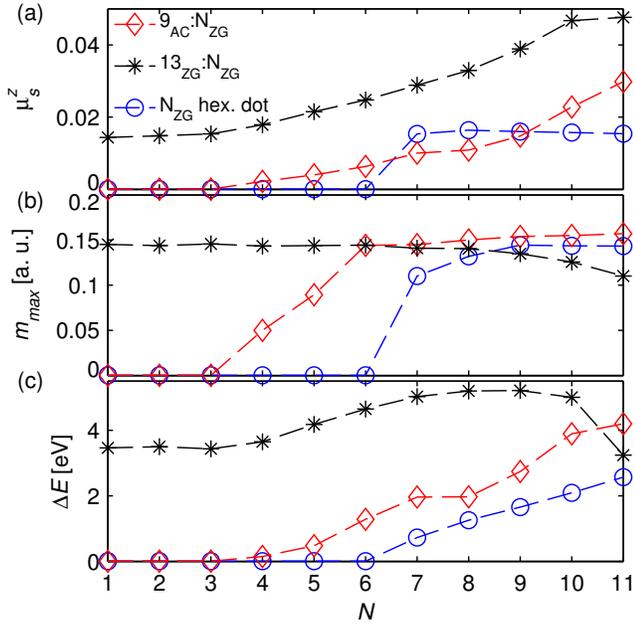}
\caption{(Color online) (a) Staggered magnetization $\mu_s^z$,
(b) maximum moment $m_{max}$ and (c) energy shift $\Delta E$
as they vary with the length of the side of the inner ring edge.}
\end{figure}

In order to demonstrate how the shape of the outer boundary
affects the distribution of the magnetic moments in the ring, we
show in Fig. 3 the magnetization in the $13_{ZG}:N_{ZG}$ rings. It
is apparent that the shape of the outer edge has a large effect
on the localization of the magnetic moments on this boundary
(compare Figs. 1(c) and 3). It is clear that in the case of zigzag
outer ring edge, the magnetization propagates much further into
the ring.

Fig. 4 displays how $\mu_s^z$, $m_{max}$, and $\Delta E$ vary with
the length of the side of the inner ring edge expressed by the
number $N$. Along with the rings whose magnetic moment
distributions were shown in Fig.~1(c) and Fig.~3, the case of a
hexagonal graphene dot having zigzag edge, is also displayed in
Fig. 4. Both the staggered magnetization $\mu_s^z$ and the energy
shift $\Delta E$ increase with $N$, i.e. with the size of the
inner ring, except for the extremely narrow $M_{ZG}:N_{ZG}$ rings.
Interestingly, the staggered magnetization in the hexagonal
quantum dots does not exceed 0.02, whereas for the
$13_{ZG}:N_{ZG}$ ring it can reach almost up to $0.05$. The nearly
twofold enlargement of the staggered magnetization could be
accounted for by the double number of zigzag edges in the
$M_{ZG}:N_{ZG}$ ring as compared to the $N_{ZG}$ graphene dot. On
the other hand, most $9_{AC}:N_{ZG}$ rings exhibit larger
staggered magnetization and all show larger maximum magnetic
moment than the hexagonal graphene dot. As a matter of fact, in
hexagonal graphene dots the zero-energy orbitals which are
localized along the adjacent zigzag sides of the edge are oriented
toward each other, whereas inner zigzag edges in rings face away
from each other. Hence, hybridization between the states of the two
edges is larger in the former case than in the latter case. This
is why $9_{AC}:N_{ZG}$ rings turn magnetic for shorter lengths of
zigzag edges than hexagonal dots (four versus seven, respectively).
The decrease of $m_{max}$ with $N$ for $13_{ZG}:N_{ZG}$ is due to
the more effective hybridization between the quasi-zero-energy
states localized on the inner and outer edges of the ring when the
ring width decreases. Hence, the electron energy shifts from the
band of zero energy states, and therefore magnetic ordering
decays, which is manifested by a smaller $m_{max}$ in the
$13_{ZG}:11_{ZG}$ ring than in the $11_{ZG}$ dot. The shapes of the
$\Delta E(N)$ curves shown in Fig. 4(c) resemble the $\mu_s^z(N)$
and $m_{max}(N)$ curves in Figs. 4(a) and 4(b).

\begin{figure}
\centering
\includegraphics[width=8.6cm]{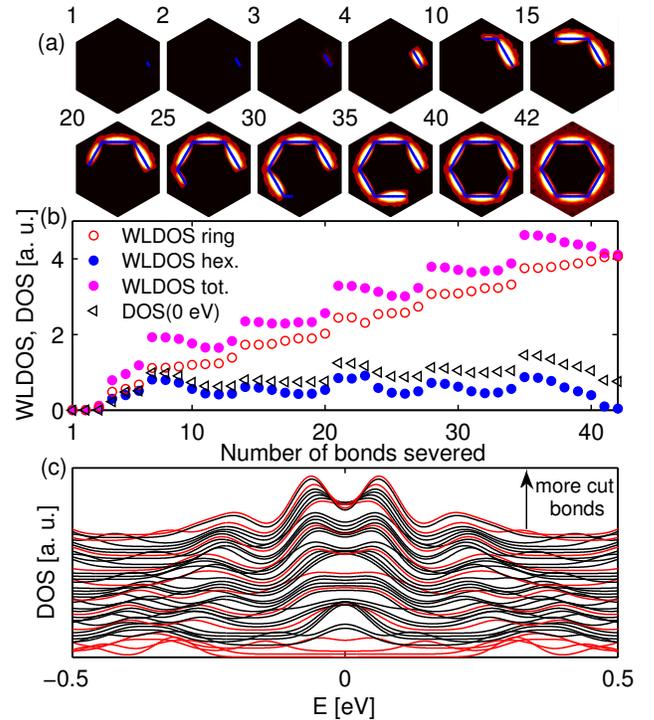}
\caption{(Color online) (a) Contour plot of the WLDOS at several
stages of the carving process forming the $7_{AC}:7_{ZG}$ ring and
the $7_{ZG}$ hexagonal dot; the number in the upper left corner
indicates the number of bonds cut. (b) Summed WLDOS in the ring,
the dot, and in the whole structure as well as the zero energy
density of states versus the number of bonds cut. (c) Stacked plot
of the density of states; red depicts the densities of the stages
displayed in panel (a).}
\end{figure}

In order to elucidate the difference between magnetic ordering in
rings and dots, one may also analyze how the local density of
states (LDOS) depends on the geometry of the structure. More
specifically, the spatial distribution of the states close to zero
energy determines how the magnetic moments evolve when the
dimensions of the structures varies. In order to enhance the
contribution of the low-energy states, we will compute the
weighted LDOS (WLDOS):~\cite{palac08}
\begin{equation}
W_{i}=\sum_j e^{-\beta E_j^2}\vert\phi_{ji}\vert^2.
\end{equation}
Here, $i$ indexes the lattice sites, $j$ labels the eigenstates,
$\beta$ is the damping coefficient chosen as $1/\sqrt\beta=0.1$
eV, whereas $\phi_{ji}$ is the value of the probability amplitude
of the $j$-th state at the site $i$. Such defined WLDOS assumes
that the contribution of the states with $\vert E_j\vert > 0.1$ eV
is negligible. The plots of the WLDOS in Fig. 5(a) illustrate how
the edge states form when the inner $7_{ZG}$ hexagonal dot is
cleaved out of the outer $7_{AC}$ hexagonal dot. The inner dot is
separated from the ring  by severing the bonds one by one between
the dot and the ring. The number of severed bonds between the dot
and the ring is explicitly shown in Fig. 5(a), and the dot edge is
depicted by the blue line. The local sublattice imbalance
accumulates quickly with the number of severed bonds, but no edge
states emerge when the number of cut bonds is less than four. The
edge states, which are depicted by red contours around the edge,
are initially distributed evenly between the ring and the dot, but they
extend more to the ring when the number of cut bonds increases.

To explore this finding in more detail, we show in Fig. 5(b) how
the total WLDOS (full purple circles), which is the sum over the
atomic sites in the dot (full blue circles) and the ring (empty
red circles), varies with the number of severed bonds. Also, the
DOS at zero energy is shown by the black triangles in Fig.~5(b).
Notice that the variation of the WLDOS has a similar shape for
each side of the ring's inner edge, and that the WLDOS displays step-like
features. These steps arise because the imbalance between the
two local sublattices, found at the ring and dot sides of the
newly formed edge, are maximized when the formation of each side
of the rings inner edge is completed. The
next side of the rings inner edge contains the opposite sublattice
imbalance, and therefore the states on this side hybridize with
the states on the previous side, which leads to a decrease of
WLDOS.~\cite{palac08} Note that after the first edge has been cut
the ring and the dot WLDOSs start deviating from each other more
strongly. This is because the hybridization in the dot is
stronger, as the edge states on adjacent segments hybridize inward
and towards each other. In the ring part the edge states face away
from each other and hybridize radially outward, hence the
hybridization is weaker. This is why the WLDOS in the former case
experiences a decline with the beginning of each new edge segment,
while in the latter case the WLDOS keeps growing. The gradual
increase of WLDOS for both cases near the end of each segment is
related to the accumulation of the local sublattice imbalances.
This pattern reappears with each new zigzag segment, with the
exception of the last bond, which after being cut results in the
separation of the two structures. By the end, the WLDOS in the
ring is much larger than WLDOS in the dot, which accounts for the
fact that the rings exhibit a larger maximum magnetic moment and
staggered magnetization than the dots. Figure 5(c) shows a stacked
plot of DOSs for each resulting structure. Plots are stacked from
the bottom up, with each subsequent line corresponding to a
structure with one more bond cut. DOSs for structures depicted in
Fig. 5(a) are colored red. It shows that only features near zero
energy evolve in a similar fashion as do the WLDOSs during the
separation of the ring and the dot. This justifies the damping of
states higher than 0.1 eV in calculating the WLDOS, as they are
not artifacts of the edge forming between the ring and the dot.

\begin{figure}
\centering
\includegraphics[width=8.6cm]{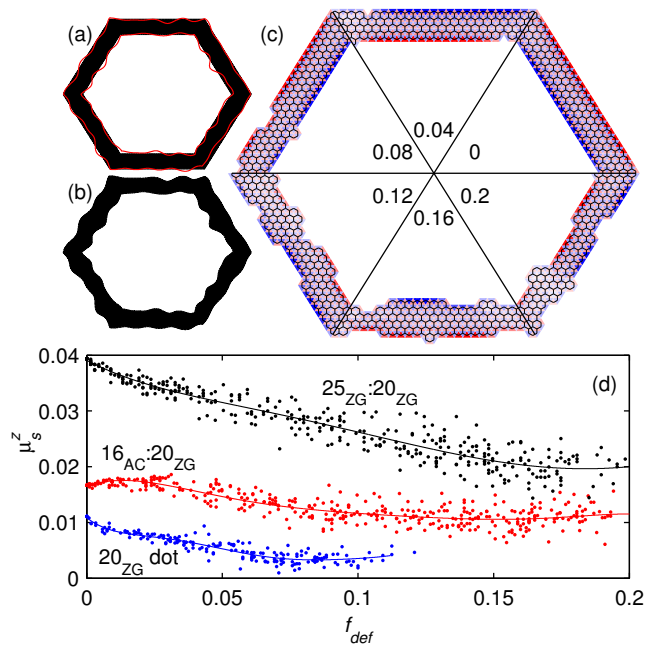}
\caption{(Color online) (a) Perfect edges (black region) are
randomly perturbed (red lines) to produce a random set of defects.
(b) Final outlook of the deformed ring. (c) Magnetic
moments distributions in $25_{ZG}:20_{ZG}$ ring for
several values of $f_{def}$. The scale is the same as in Fig.~3.
(d) Staggered magnetization of an ensemble of randomly
defected structures as a function of defect ratio for
$25_{ZG}:20_{ZG}$ ring (black dots), $16_{AC}:20_{ZG}$ ring
(red dots) ring, and $20_{ZG}$ hexagonal dot (blue dots). The
polynomial fitting curves are added to guide the eye.}
\end{figure}

Finally, we examine the influence of the edge deformations on
somewhat larger structures; namely, the $25_{ZG}:20_{ZG}$ and
$16_{AC}:20_{ZG}$ rings and the $20_{ZG}$ dot. Larger structures are
considered here because they can be deformed in a larger number of ways
than smaller structures analyzed in the rest of the paper. Defects are induced
by randomly deforming the polygons which outline the perfect
structure as shown in Fig.~6 (a). The amplitude of this
deformation is itself a randomly selected number out of a specific
range and the final structure is made up of all atoms that are
enclosed by the deformed outline,\cite{futnota2} which is shown in
Fig.~6(b). In order to quantify the amount of defects, the defect
ratio $f_{def}$ is defined as a fraction of the total number of
the defects, which is a sum of the missing and the surplus sites,
and the number of the sites in the original structure. The
magnetic moment distributions in the $25_{ZG}:20_{ZG}$ ring for a
few values of $f_{def}$ are shown in Fig.~6 (c). Also, variation
of the staggered magnetization with the defect fraction for the
$16_{AC}:20_{ZG}$ and $25_{ZG}:20_{ZG}$ rings and the $20_{ZG}$
dot is displayed in Fig.~6 (d). For the $25_{ZG}:20_{ZG}$ ring and
the $20_{ZG}$ hexagon, $\mu_s^z$ decreases with defect fraction. This
is expected, having in mind that the defects can only impair the
conditions for magnetism in zigzag edges. On the other hand, for the $16_{AC}:20_{ZG}$
ring, small random defects are more likely to make the larger outer
edge magnetic than to make the smaller inner edge nonmagnetic.
This explains the initial rise in $\mu_s^z$ for $f_{def}$ up to 0.02.

In conclusion, we predict an antiferromagnetic phase in hexagonal
graphene rings with zigzag inner edge within the mean-field
Hubbard model. The distribution of magnetic moments is found to
strongly depend on the type of outer edge, and larger
antiferromagnetic order is found in rings than in hexagonal dots.
Peculiar hybridization between the states of adjacent sides of the
inner ring edge is found to lead to an increase of magnetization
of rings with respect to dots. Also, the staggered magnetization
and the maximum magnetic moment are found to be strongly
influenced by the size and the shape of the rings. For wide rings,
the maximum magnetic moment is largest when both the inner and
outer edges are zigzag. But, as a consequence of the hybridization
between the states of the two edges, the maximum magnetic moment
in a ring with armchair outer edge exceeds the one for the zigzag
outer edge when the ring width decreases. The staggered
magnetization in both the hexagonal dots and the rings with zigzag
outer edge is found to decrease faster than in the rings with
armchair outer edge when the number of the edge defects increases.

This work was supported by the EuroGRAPHENE programme of the ESF
(project CONGRAN), the Serbian Ministry of Education, Science, and
Technological Development, and the Flemish Science Foundation
(FWO-Vl).


\begin{thebibliography}{00}
\bibitem{futnota1}We also include the constant part $U/4-U/2\sum_{i,\sigma}n_{i,\sigma}$ in the Hamiltonian.
\bibitem{futnota2}We remove all the dangling atoms which might appear after the edge is deformed.
Also, we ensure that all the deformed structures have equal
numbers of the A and B sublattice sites, thus our calculations
always regarded the antiferromagnetic limit.
\end{thebibliography}
\end{document}